\documentclass[aps,twocolumn,showpacs,floatfix]{revtex4}
\usepackage{epsfig}
\usepackage{graphicx}
\usepackage{dcolumn}
\usepackage{amsthm,amsmath}

\begin{document}
\title{Wave packet construction in three-dimensional quantum billiards: Visualizing the closed orbit, collapse and revival of wave packets in the cubical billiard }
\author{Maninder Kaur\footnote{Email: mannu\_711@yahoo.co.in}, Bindiya Arora\footnote{Email: arorabindiya@gmail.com} and M. Mian}
\affiliation{Department of Physics, Guru Nanak Dev University, Amritsar, Punjab-143005, India}
\date{Received date; Accepted date}

\begin{abstract}
We examine the dynamical evolution of wave packets in a cubical billiard where three quantum numbers ($n_x,n_y,n_z$) determine its energy spectrum and consequently its dynamical behavior. We have constructed the wave packet  in the cubical billiard  and have observed its time evolution for various closed orbits. The closed orbits are possible for certain specific values of quantum numbers ($n_x,n_y,n_z$) and initial momenta ($k_x,k_y,k_z$). We  observe that a cubical billiard exhibits degenerate energy levels and the path lengths of the closed orbits for these degenerate energy levels are  identical. In spite of the identical path lengths, the shapes of the closed orbits for degenerate levels are different and depend upon angles `$\theta$' and `$\phi$’' which we term as the sweep  and the elevation angle respectively. These degenerate levels owe their origin to the symmetries prevailing in the cubical billiard and degenerate levels disappear completely or partially for a parallelepiped billiard as the symmetry  breaks  due to commensurate or incommensurate ratio of sides.
\end{abstract}

\maketitle
\section{Introduction}~\label{sec1}
The study of the time evolution of the unbound and bound state wave packet illuminates many features of the wave mechanics. This includes both semi-classical features as well as purely quantum mechanical effects. In the classical regime the wave packet mimics a classical particle and in the quantum regime the wave packet undergoes a sequence of collapses and revivals as a manifestation of the underlying quantum interference effects. The dynamics of the unbound and bound wave packet have been visualized by many authors~\cite{10,11,12,13,u1,h1,h2,h3,h4}. This approach of visualization of the dynamics of the wave packet has facilitated the understanding of highly nonclassical and the least intuitive quantum phenomenon. As a consequence such visualizations have become increasingly relevant as pedagogical tool.	
The dynamical behavior of the bound state wave packets in the quantum systems where the energy spectrum depends on a single quantum number has been discussed by several authors~\cite{1,2,3,4,b1}. In all these bound state quantum systems, the  dynamics of the wave packets (which are mostly the Gaussian wave packets) lucidly reveal the existence of the phenomenon of the collapses, revivals, fractional revivals and the super revivals of the wave packets. Moreover, the study of the dynamics of the initial Gaussian wave packet (GWP) trapped inside bound potential wells can provide information regarding the shape and length of the closed orbits. For a system with only one degree of freedom the closed orbits are simple to and fro trajectories with length equal to double the width of  bounding potential. The system has been widely studied by a number of groups in two-dimensions~\cite{5,7,8,t1}. The motivation to extend the study to a three-dimensional system is the existence of real world systems that can not be reduced to $2D$ idealization and exhibits some (both classical and quantum) genuine 3D effects.

Billiards are widely being studied in the field of quantum chaos~\cite{14} and play an important role in extending this field to the system with more than two degrees of freedom~\cite{15,16,17,18}. They are also central to the investigation of three dimensional chaos in the resonant optical cavities~\cite{19}. In  theory of quantum chaos, connection between the chaotic behavior of classical systems and specific properties of their quantum mechanical counterparts has received much attention~\cite{m1,m2,m5}. One way to make this connection is to look at the eigenstate of billiards which exhibit a structure corresponding to the classical orbits~\cite{m9}. These structures, called scars have been directly related to unstable periodic orbits~\cite{m12,m14} . Visual inspection of movies that show the time evolution of the probability distribution of a quantum particle moving in two and three dimensional billiard of different shapes strongly suggest that these sequence of images may contain enough information to distinguish between classical, integrable and chaotic systems~\cite{20}. The other important application of theory of quantum billiards is related to the double-clad fibres. In such a fibre, a small core with low numerical aperture confines the signal and the wide cladding confines the multi-mode pump. In the paraxial approximation, the complex field of pump in the cladding behaves like a wave function in  quantum billiard~\cite{21}.

In this work, we study the billiard problem in 3-dimensions. We have observed that in a cubical billiard, the degenerate energy levels belong to those set of quantum numbers ($n_x,n_y,n_z$) for which the expression $n_x^2+n_y^2+n_z^2 = D$ evaluates to an identical constant. Further for a cubical billiard the path lengths of closed orbits are proportional to the quantity $\sqrt{n_x^2+n_y^2+n_z^2}$. Hence the lengths of the closed orbits for the degenerate energy levels are equal. We have observed that in spite of the identical path lengths, the shapes of the closed orbits for these degenerate levels are different and depend upon two angles `$\theta$' and `‘$\phi$’' which we term as the sweep angle and the elevation angle, respectively.

We begin in  section \ref{gwp}, by discussing the basic properties of a Gaussian wave packet. Further we review the various time scales viz. classical time period, revival time period and the super revival time period for a general one dimensional bound system. In section \ref{isw},  we employ the general formulae of section \ref{gwp}, to completely define the specific bound system of 1D infinite square well (ISW) centered at x=0. In section \ref{cb}, we formulate the theoretical background for a cubical billiard. We discuss the general time dependence, condition for the closed orbits and the wave packet construction in the cubical billiard. Here we also define $\theta$ and $\phi$. In the same section, we discuss the degenerate states in the cubical billiard and the shapes of the closed orbits for the set of degenerate states. In section \ref{pb}, we  provide a general overview of the results for the case of a parallelepiped billiard. In section~\ref{rd}, we present our results for the dynamics of the GWP along various closed orbits in a cubical billiard and the pattern of collapses and revivals of the GWP .The results have been presented in the form of the plots of position probability density at sequential instants of time within one classical time period and auto-correlation function for the short time (time $\sim$ few classical time periods) as well as the long time  (time $\sim$ few revival time periods). We have also produced the animations which clearly illustrate the dynamics of the wave packet and shapes of the various closed orbits depending upon the different initial conditions.

\section{Gaussian wave packet}\label{gwp}
 Consider a Gaussian wave packet (GWP) described by the wave function
\begin{equation}
\psi_(x,0) = \frac{1}{\sqrt{\sigma_0\sqrt{\pi}}}\rm{exp}^{\iota k_0(x-x_0)}\rm{exp}^{-(x-x_0)^2/2\sigma_0^2}.\label{gwe}
\end{equation}
This GWP is centered at $x=x_0$, with standard deviation, $ \frac{\sigma_0}{\sqrt{2}}$ .
The corresponding momentum space wave function is
\begin{equation}
\psi(k,0) = \frac{1}{\sqrt{\sigma_0\sqrt{\pi}}}\rm{exp}^{-\sigma_0^2(k-k_0)^2/2}\rm{exp}^{\iota k_0(k-k_0)}.
\end{equation}
This GWP is centered at $k=k_0$, with standard deviation  $ \frac{1}{\sqrt{2}\sigma_0}$.
Here $x_0$ and $k_0$  correspond to initial position and  initial momentum  of the wave packet.
The expectation value  of the total kinetic energy is given by
\begin{equation}
\left<\hat{E}\right>=\frac{1}{2m}\left<k^2\right>\left(k_0^2+\frac{1}{2\sigma_0^2}\right) \rm{with}
  <k^2> = k_0^2+2\sigma_0^2. \nonumber
\end{equation}
If we expand such an initial  GWP in terms of the energy eigen states $U_n(x)$ of a general one dimensional bound state potential~\cite{7}, we have
\begin{equation}
\psi(x,0) = \sum_{n=1}^{\infty}a_n U_n(x),
\end{equation}
where
\begin{equation}
a_n=\int_{-\infty}^{\infty}U_n^{*}(x)\psi(x,0)dx.\label{eq-an}
\end{equation}
The resulting $a_n$ must satisfy the two constraints, namely
\begin{eqnarray}
\sum_{n=1}^{\infty}\left|a_n^2\right|&=&0, \\
\sum_{n=1}^{\infty}\left|a_n^2\right|E_n&=&\left<\hat{E}\right>,
\end{eqnarray}
where $E_n$ represents the energy eigen values of the bound state potential. The time dependence of the initial GWP as given by the solution of the time dependent Schr$\ddot{\rm{o}}$dinger wave equation is
\begin{equation}
\psi(x,t)=\sum_{n=1}^{\infty}a_nU_n(x)\rm{exp}(-\iota E_nt/\hbar),\label{time}
\end{equation}
The coefficients $a_n$ are defined by Eq.~\ref{eq-an}. By taking the summation over $n$  we restrict ourselves to the superposition of bound states only. This senario appears when a GWP is trapped inside some bound state potential or a localized wave packet is produced using a short pulse laser experimentally ~\cite{l1,l2}. Physically this localization is equivalent to transformation of the Gaussian wave packet from the x-space to the n-space via Eq.~\ref{eq-an}. Thus we will end up with a Gaussian wave packet centered around some central value of the principle quantum number ‘$n_0$’ instead of a Gaussian wave packet centered around some position coordinate $x_0$’. The mapping from the x-space to the n-space is brought about by the expansion coefficient $a_n$’. Thus it follows that the expansion is centered around a mean value $n_0$. If we expand the energy in a Taylor series around this mean value $n_0$, we get\cite{5}
\begin{equation}
E_n\approx E_{n_{0}} + (n-n_0)E_{n_0}' +\frac{1}{2}(n-n_0)^2E_{n_0}''+\frac{1}{6}(n-n_0)^3E_{n_0}'''+.....,
\end{equation}
where the primes over the energy terms represent derivatives of first order, second order, third order and so on.
These derivatives define the various time scales depending upon the mean value $n_0$. The first time scale
\begin{equation}
T_{cl} = \frac{2\pi}{E'_{n_0}}
\end{equation}
is the classical time period for the shortest closed orbit. It controls the initial behavior of the wave packet. The second time scale
\begin{equation}
T_{rev}=\frac{2\pi}{\frac{1}{2}E''_{n_0}}
\end{equation}
is the revival  time. It governs the appearance of the fractional  revivals and the full revivals. The third time scale
\begin{equation}
T_{sr} = \frac{2\pi}{\frac{1}{6}E'''_{n_0}}
\end{equation}
is the super revival time. It is a larger time scale as compared to the classical time period and the revival time period .

The short term  behavior  and the long term behavior of the wave packet is analyzed  using a standard tool popularly known as the auto correlation function ~\cite{6}, given as
\begin{eqnarray}
A(t)&=&\int_{-\infty}^{\infty}\psi*(x,t)\psi(x,0)dx \\
&=&\int_{-\infty}^{\infty}\phi*(k,t)\phi(k,0)dt \\
&=& \sum_{n=1}^{\infty}\left|a_n\right|^2\rm{exp}(\iota E_n\hbar).
\end{eqnarray}
This measures the degree of overlap of the initial wave function with itself at later times.

\section{One Dimensional Infinite Square Well}\label{isw}
We  consider a one dimensional infinite square well centered at $x=0$ and defined by,
\begin{eqnarray}
V(x) &=& 0\mbox{  }\rm{for}\mbox{  }  -1<x<1, \nonumber\\
     &=& \infty \mbox{  }\rm{for} \mbox{  }   |x|>1.\nonumber
\end{eqnarray}
The eigen function and the eigen values of this quantum system are well defined and are given by (here $2m=\hbar=1$ ),
\begin{eqnarray}
U_n(x)&=&\sin (k_n(1+x)), \\
E_n&=&k_n^2,\label{18} \\
k_n&=&\frac{n\pi}{\rm{L}}=\frac{n\pi}{2}.
\end{eqnarray}
For  a GWP defined by Eq.~\ref{gwe} confined within this infinite square well the expansion co-efficient are evaluated using~\cite{7},
\begin{eqnarray}
a_n&=&\int_{-\infty}^{\infty}U^*_n(x)\psi(x,0)dx, \nonumber\\
&=&\int_{-\infty}^{\infty}\sin(k_n(1+x))\psi(x,0)dx,\nonumber
\end{eqnarray}
\begin{eqnarray}
=\frac{1}{\sqrt{\sigma_0\sqrt{\pi}}}\int_{-\infty}^{\infty}\sin(k_n(1+x))\exp\left(\iota k_0(x-x_0\right))\nonumber\\
\exp\left(-\frac{(x-x_0)^2}{2\sigma_0^2}\right)dx.
\end{eqnarray}
Using the standard integral
\begin{equation}
\int_{-\infty}^{\infty}\exp(-\alpha x^2-bx)dx=\sqrt{\frac{\pi}{\alpha}}\exp{\left(\frac{\beta^2}{4\alpha}\right)},
\end{equation}
and
\begin{equation}
\sin(k_n(1+x)) = \frac{\exp{\iota k_n(1+x)}-\exp{-\iota k_n(1+x)}}{2\iota},
\end{equation}
we get
\begin{eqnarray}
a_n=\frac{\sqrt{2\sigma_0\sqrt{\pi}}}{2\iota}[e^{\iota (k_n(1+x_0)}e^{-\frac{\sigma_0^2}{8}(k_0+n\pi)^2} \nonumber\\
-e^{-\iota (k_n(1+x_0)}e^{-\frac{\sigma_0^2}{8}(k_0-n\pi)^2} ].
\end{eqnarray}
In passing, we state that even though the GWP is almost normalized, yet we ensure complete normalization of the GWP with  the help of analytic calculations on $a_n$.
The closed orbits of this GWP in the $1D$ ISW are simple to and fro trajectories. The  classical time period and the revival time period  are given by
\begin{eqnarray}
T_{cl} &=& \frac{2\pi}{\left|E'_{n_0}\right|} = \frac{2\pi}{\pi^2n_0/2}=\frac{4}{n_0\pi}, \\
T_{rev}&=& \frac{2\pi}{\frac{1}{2}\left|E''_{n_0}\right|} = \frac{2\pi}{\pi^2/4}=\frac{8}{\pi}.
\end{eqnarray}
where $E_n$ is defined by Eq.~\ref{18}

\section{Cubical Billiard}\label{cb}
For a cubical billiard i.e. $3D$ ISW with dimensions $L_x \times L_y \times L_z = a \times a \times a$, the problem is equivalent to three $1D$ ISWs along the x, y and z directions and the Eigen functions and the Eigen values are given as
\begin{eqnarray}
w(x,y,z)&=&U_{n_x}(x)U_{n_y}(y)U_{n_z}(z)\\
E_{n_x,n_y,n_z}&=&k_{n_x}^2+k_{n_y}^2+k_{n_z}^2,\label{eq-energy}
\end{eqnarray}
where $n_x$, $n_y$, and $n_z$ are appropriate quantum numbers and
\begin{eqnarray}
k_{n_x}&=&\frac{n_x\pi}{\rm{L_x}}=\frac{n_x\pi}{2}\nonumber\\
k_{n_y}&=&\frac{n_y\pi}{\rm{L_y}}=\frac{n_y\pi}{2}\nonumber\\
k_{n_z}&=&\frac{n_z\pi}{\rm{L_z}}=\frac{n_z\pi}{2}\nonumber\\
U_{n_x}(x) &= &\sin(k_{n_x}(1+x)) \nonumber\\
U_{n_y}(y) &= &\sin(k_{n_y}(1+y)) \nonumber\\
U_{n_z}(z)& = &\sin(k_{n_z}(1+z)) \nonumber
\end{eqnarray}

\subsection{Closed orbits }
Inside a cubical billiard the particle is free so that its total kinetic energy is constant implying that $v_x^2+v_y^2+v_z^2$
is fixed. In fact the velocity components along the respective axis  i.e. $v_x, v_y$ and $v_z$ are separately constant.The resultant magnitude and  resultant direction of these velocity components follows from the laws of vector addition. If $v_n$  represents the resultant velocity and $k_n=mv_n$ represents the resultant momentum,  the GWP proceeds along that resultant direction with  net momentum $k_n $. This gives rise to a particular trajectory for a specific combination of $v_x ,v_y $ and $v_z$. With change in any of the component velocity, the resultant trajectory alters. But all these possible trajectories do not result into a closed orbit. The closed orbits exist only if the following conditions are satisfied~\cite{7}
$$\frac{k_{0y}}{k_{0x}}=\frac{n_y}{n_x};\frac{k_{0z}}{k_{0x}}=\frac{n_z}{n_x};\frac{k_{0z}}{k_{0y}}=\frac{n_z}{n_y}$$.
If $p, q, r$ are numbers such that $2p, 2q, 2r$ give the number of hits on walls along the $x, y, z$ directions respectively, then in terms of these numbers the condition for the closed orbits is given by
\begin{eqnarray}
n_x:n_y:n_z = p:q:r=\rm{ a\mbox{ }rational\mbox{ }number}.\label{eq-rational}
\end{eqnarray}
The length of the closed orbit L$(p,q,r)$ can be obtained from the simple geometrical construction~\cite{9} and is given by
\begin{equation}
{\rm L}(p,q,r) = 2a\sqrt{p^2+q^2+r^2}.
\end{equation}
The time period for the closed orbit is given by
\begin{eqnarray}
T_{cl}^{(p_0)}&=& \frac{{\rm L}(p,q,r)}{v_0},\label{tpo} \\
&=&\tau_{cl}\sqrt{p^2+q^2+r^2},\nonumber\\
\frac{T_{cl}^{(p_0)}}{\tau_{cl}}&=&\sqrt{p^2+q^2+r^2}.
\end{eqnarray}
Here $v_0$ is the classical speed and $\tau_{cl}=\frac{2a}{v_0}$ is the classical time period for the simplest to and fro motion.
Now,
\begin{equation}
T_{rev}^{(p_0)}=\frac{2\pi}{\frac{1}{2}|E_n''|},
\end{equation}
with $E_n=\frac{n^2\pi^2}{L(p,q,r)}$ which gives $E_n''=\frac{2\pi^2}{L^2(p,q,r)}$. Subsitituing for $E_n''$ for revival time we get
\begin{eqnarray}
T_{rev}^{(p_0)}&=&\frac{2}{\pi}L(p,q,r)\nonumber\\
&=&\frac{8}{\pi}a^2(p^2+q^2+r^2).\nonumber
\end{eqnarray}
The above equation is simplified to a more useful form written as follows
\begin{equation}
\frac{T_{rev}^{(p_0)}}{\tau_{rev}}=p^2+q^2+r^2,
\end{equation}
where $\tau_{rev}=\frac{8a^2}{\pi}$ is the revival time period for the shortest to and fro closed orbit.

Next, we calculate the values of $n_x,n_y$  and $n_z$ which  will lead to the closed orbits for the motion of the wave function. The evaluation of the quantum numbers $n_x$, $n_y$ and $n_z$ is based upon the total energy condition (eq.~\ref{eq-energy}) i.e.
\begin{eqnarray}
E(n_x,n_y,n_z)&=& \frac{1}{2}mv_n^2,\nonumber\\
				&=&\frac{\pi^2}{4}\left(\frac{p^2+q^2+r^2}{p^2}\right)\times n_x^2.
\end{eqnarray}
From the above equation we get
\begin{equation}
n_x = \frac{v_np}{\pi\sqrt{p^2+q^2+r^2}} \\
\end{equation}
Similarly,\\
\begin{eqnarray}
n_y &=& \frac{v_nq}{\pi\sqrt{p^2+q^2+r^2}} \\
n_z &=& \frac{v_nr}{\pi\sqrt{p^2+q^2+r^2}}.
\end{eqnarray}

From Eq.~\ref{eq-rational} it follows that a particular closed orbit comes into picture for an appropriate choice of $n_x$, $n_y$ and $n_z$. Consider such a closed orbit as shown in Fig.~\ref {fig-1}. The construction of this closed orbit requires the knowledge of the sweep angle $\theta$, elevation angle $\phi$ and the momentum components along the x, y, z axis. All these factors are defined below
\begin{eqnarray}
\tan\theta &=& \frac{q}{p}\label{th}\\
\tan\phi &=& \frac{r}{\sqrt{p^2+q^2}}\label{phi}\\
k_{0x} &=& k_0\frac{\sqrt{p^2+q^2}}{\sqrt{p^2+q^2+r^2}}\cos\theta \\
k_{0y} &=& k_0\frac{\sqrt{p^2+q^2}}{\sqrt{p^2+q^2+r^2}}\sin\theta \\
k_{0z} &=& k_0\sin\phi.
\end{eqnarray}
\subsection{Wave packet construction}
Time evolution of the initial GWP inside an ISW with three degrees of freedom is given by the following equation
\begin{equation}
\phi(x,y,z)  = \psi(x,t)\psi(y,t)\psi(z,t),
\end{equation}
where
\begin{eqnarray}
\psi(x,t) &=&\sum_{n_x=1}^{\infty}a_{n_x}U_{n_x}(x)\exp(-\iota E(n_x)t/\hbar),\nonumber \\
\psi(y,t) &=&\sum_{n_y=1}^{\infty}a_{n_y}U_{n_y}(y)\exp(-\iota E(n_y)t/\hbar), \nonumber\\
\psi(z,t) &=&\sum_{n_z=1}^{\infty}a_{n_z}U_{n_z}(z)\exp(-\iota E(n_z)t/\hbar).\nonumber
\end{eqnarray}
Here,
\begin{eqnarray}
a_{n_x} = \frac{\sqrt{2\sigma_0\sqrt{\pi}}}{2\iota}[e^{\iota (k_{n_x}(1+x_0)}e^{-\frac{\sigma_0^2}{8}(k_{0x}+n_x\pi)^2}\nonumber \\
-e^{-\iota (k_{n_x}(1+x_0)}e^{-\frac{\sigma_0^2}{8}(k_{0x}-n_x\pi)^2} ],\nonumber \\
a_{n_y} = \frac{\sqrt{2\sigma_0\sqrt{\pi}}}{2\iota}[e^{\iota (k_{n_y}(1+y_0)}e^{-\frac{\sigma_0^2}{8}(k_{0y}+n_y\pi)^2} \nonumber\\
-e^{-\iota (k_{n_y}(1+y_0)}e^{-\frac{\sigma_0^2}{8}(k_{0y}-n_y\pi)^2} ], \nonumber\\
\rm{and}\\
a_{n_z} = \frac{\sqrt{2\sigma_0\sqrt{\pi}}}{2\iota}[e^{\iota (k_{n_z}(1+z_0)}e^{-\frac{\sigma_0^2}{8}(k_{0z}+n_z\pi)^2}\nonumber \\
-e^{-\iota (k_{n_z}(1+z_0)}e^{-\frac{\sigma_0^2}{8}(k_{0z}-n_z\pi)^2}]. \nonumber
\end{eqnarray}

\subsection{Degenerate states and closed orbits}\label{ds}
The connection between the classical path length and the energy spectrum is well known i.e. the Fourier transform of the density of quantized energy levels  exhibit a delta function like peaks at distance values ($L$) corresponding to the length of closed orbits~\cite{9}. Thus, it is obvious that the various degenerate energy levels will correspond to the identical path lengths of the closed orbits.
However the shapes of the closed orbits corresponding to degenerate energy levels depends upon the sweep angle `$\theta$' and the elevation angle `$\phi$', which can not be obtained from the Fourier analysis. Both these angles depend upon $p, q, r$ or $n_x$, $n_y$, $n_z$ and have been calculated in this work using Eq.~\ref{th} and Eq.~\ref{phi}. Table~\ref{tab1} enlists some of the possible  values of the sweep angle $\theta$, elevation angle $\phi$, time period of the corresponding closed orbit and the degeneracy factor for the various combinations of integral values of $p, q, r$.

\begin{table}[h!]\label{tab1}
\caption{\label{tab1}  Some of the possible values of the sweep angle `$\theta$', elevation angle `$\phi$', ratio of classical time period and classical time period of the shortest closed orbit $T_{cl}^{po}/\tau_{cl}$, ratio of revival time period and revival time period of the shortest closed orbit $T_{rev}^{po}/\tau_{rev}$  and the degeneracy factor for the cubical billiard characterized by the integral values of  $p, q, r$.}
\begin{ruledtabular}
\begin{tabular}{cccccc}
($p,q,r$) & $\theta$ & $\phi$ & $T_{cl}^{po}/\tau_{cl}$ & $T_{rv}^{po}/\tau_{rv}$ & Degeneracy \\
(1,1,1) &  $45^\circ$ & $35.26^\circ$ & $\sqrt{3}$ & 3 & 1 \\
(1,1,2) &  $45^\circ$ & $54.74^\circ$ & $\sqrt{6}$ & 6 & \\
(1,2,1) &  $63.44^\circ$ & $24.09^\circ$ & $\sqrt{6}$ & 6 & 3 \\
(2,1,1) &  $26.57^\circ$ & $24.09^\circ$ & $\sqrt{6}$ &  6 & \\
(1,1,3) &  $45^\circ$ & $64.76^\circ$ & $\sqrt{11}$ &  11 & \\
(1,3,1) &  $71.57^\circ$ & $17.55^\circ$ & $\sqrt{11}$ &  11 &3 \\
(3,1,1) &  $18.43^\circ$ & $17.55^\circ$ & $\sqrt{11}$ &   11 & \\
(1,2,3) &  $63.44^\circ$ & $53.30^\circ$ & $\sqrt{14}$ &  14 & \\
(3,1,2) &  $18.43^\circ$ & $32.31^\circ$ & $\sqrt{14}$ &   14 & \\
(2,3,1) &  $56.30^\circ$ & $15.50^\circ$ & $\sqrt{14}$ &   14 & 6 \\
(1,3,2) &  $71.57^\circ$ & $32.31^\circ$ & $\sqrt{14}$ &   14 & \\
(3,2,1) &  $33.69^\circ$ & $15.50^\circ$ & $\sqrt{14}$ &   14 & \\
(2,1,3) &  $26.57^\circ$ & $53.30^\circ$ & $\sqrt{14}$ &  14 &
\end{tabular}
\end{ruledtabular}
\end{table}


\section{Parallelepiped billiard}\label{pb}
Consider  a parallelepiped billiard with sides $L_x \times L_y \times L_z=L \times2L \times 4L$.  For this case the condition for closed orbits can be obtained from the fact that the classical close or periodic orbits are reproduced when the classical periods corresponding to the three quantum numbers $n_x,n_y,n_z $ are commensurate~\cite{7}, i.e.
\begin{eqnarray}
pT_{cl}^{(n_x)}&=&qT_{cl}^{(n_y)}=rT_{cl}^{(n_z)} =T_{cl}^{(po)},\nonumber \\
\frac{p}{\pi}\frac{L_x^2}{n_x}&=&\frac{q}{\pi}\frac{L_y^2}{n_y}=\frac{r}{\pi}\frac{L_z^2}{n_z},\nonumber\\
\frac{pL_x}{\frac{n_x}{L_x}}&=&\frac{pL_y}{\frac{n_y}{L_y}}=\frac{pL_z}{\frac{n_z}{L_z}}.
\end{eqnarray}
Thus, the condition for the closed orbit is modified to
\begin{equation}
\frac{n_x}{L_x}:\frac{n_y}{L_y}:\frac{n_z}{L_z}=pL_x:qL_y:rL_z,\label{eq-ratio}
\end{equation}
and the total energy condition becomes
\begin{equation}
E=\frac{1}{2}mv_0^2=\pi^2\left(\frac{n_x^2}{L_x^2}+\frac{n_y^2}{L_y^2}+\frac{n_z^2}{L_z^2}\right).\label{eq-en}
\end{equation}
Equations (\ref{eq-ratio}) and (\ref{eq-en}) lead to
\begin{eqnarray}
\frac{n_x}{L_x}&=&\frac{v_o}{2\pi}\left[\frac{pL_x}{\sqrt{(pL_x)^2+(qL_y)^2+(rL_z)^2}}\right],\nonumber\\
\frac{n_y}{L_y}&=&\frac{v_o}{2\pi}\left[\frac{qL_y}{\sqrt{(pL_x)^2+(qL_y)^2+(rL_z)^2}}\right],\nonumber\\
\frac{n_z}{L_z}&=&\frac{v_o}{2\pi}\left[\frac{rL_z}{\sqrt{(pL_x)^2+(qL_y)^2+(rL_z)^2}}\right].
\end{eqnarray}
The length of the closed orbit is given by
\begin{eqnarray}
L(pL_x,qL_y,rL_z)&=&2\sqrt{(pL_x)^2+(qL_y)^2+(rL_z)^2},\label{l} \nonumber\\
                &=&2\sqrt{(p_1)^2+(q_1)^2+(r_1)^2}.
\end{eqnarray}
where $pL_x=p_1$, $qL_y=q_1$, and $rL_z=r_1$.
For a parallelepiped billiard with sides $L_x\times L_y \times L_z=L \times 2L\times 4L$, the shortest closed orbit is along the x-axis, so we define the time period of the closed orbits with respect to the classical time period of the shortest to and fro trajectory along the x-axis. The classical time period of the shortest to and fro trajectory along this axis is defined as  $\tau_x=\frac{2L_x}{v_{0x}}$. From equations (~\ref{tpo}) and (~\ref{l}), we can deduce

\begin{equation}
T_{cl}^{(p_0)}(n_x) = \frac{2L(p,q,r)}{v_{0x}}=\frac{2\sqrt{(pL_x)^2+(qL_y)^2+(rL_z)^2}}{v_{0x}},
\end{equation}
Since, $L_x:L_y:L_z=1:2:4$,
\begin{eqnarray}
T_{cl}^{(p_0)}(n_x) &=&\frac{2L_x\sqrt{p^2+(2q)^2+(4r)^2}}{v_{0x}},\nonumber\\
\frac{T_{cl}^{(p_0)}(n_x)}{\tau_x}&=& \sqrt{p^2+(2q)^2+(4r)^2}.
\end{eqnarray}
Also the expressions  for $\tan(\theta)$  and $\tan(\phi)$  modify as follows,
\begin{eqnarray}
\tan(\theta)&=&\frac{q_1}{p_1}=\frac{qL_x}{pL_y},\nonumber\\
\tan(\phi)&=&\frac{r}{\sqrt{p_1^2+q_1^2}}=\frac{rL_z}{\sqrt{(pL_x)^2+(qL_y)^2}},
\end{eqnarray}
And the condition for the degeneracy modifies to,
\begin{equation}
\left(\frac{n_x}{L_x}\right)^2+\left(\frac{n_y}{L_y}\right)^2+\left(\frac{n_z}{L_z}\right)^2=D,
\end{equation}
where D is a constant. Table~\ref{tab2} enlists  values of the sweep angle $\theta$, elevation angle $\phi$, time period of the corresponding closed orbit and the degeneracy factor for some of the possible combinations of integral values of $p, q, r$ for the parallelepiped billiard.

\begin{table}[h!]
\caption{\label{tab2}  Values of the sweep angle `$\theta$', elevation angle `$\phi$', ratio of classical time period and classical time period of the shortest closed orbit $T_{cl}^{po}/\tau_{cl}$, ratio of revival time period and revival time period of the shortest closed orbit $T_{rev}^{po}/\tau_{rev}$ and the degenracy factor for some of the possible combinations of integral values of $p, q, r$ for the parallelepiped billiard.}
\begin{ruledtabular}
\begin{tabular}{cccccc}
($p,q,r$) & $\theta$ & $\phi$ & $T_{cl}^{po}/\tau_{cl}$ &  $T_{rv}^{po}/\tau_{rv}$ & Degeneracy\\
(1,0,0) &  $0^\circ$ & $0\circ$ & $1$ & 1 & 1 \\
(0,1,0) &  $90^\circ$ & $0^\circ$ & $2$ & 1 &  1\\
(0,0,1) &  $90^\circ$ & $90^\circ$ & $4$ & 1 &  1 \\
(1,1,1) &  $63.44^\circ$ & $60.79^\circ$ & $\sqrt{21}$ & 21 & 1 \\
(1,1,2) &  $63.44^\circ$ & $74.38^\circ$ & $\sqrt{69}$ & 69 & 1 \\
(1,2,1) &  $75.96^\circ$ & $44.13^\circ$ & $\sqrt{33}$ & 33 & 1 \\
(2,1,1) &  $45^\circ$ & $54.74^\circ$ & $\sqrt{24}$ & 24 & 1  \\
(3,1,1) &  $33.69^\circ$ & $47.97^\circ$ & $\sqrt{29}$ & 29 & 1 \\
(1,1,3) &  $63.43^\circ$ & $79.44^\circ$ & $\sqrt{149}$ &149&  1 \\
(1,3,1) &  $80.54^\circ$ & $33.33^\circ$ & $\sqrt{53}$ & 53 & 1 \\
(2,2,1) &  $63.43^\circ$ & $41.81^\circ$ & $\sqrt{36}$ & 36 &  \\
(4,1,1) &  $26.57^\circ$ & $41.81^\circ$ & $\sqrt{36}$ &36 &  2 \\
(2,3,1) &  $71.57^\circ$ & $32.31^\circ$ & $\sqrt{56}$ &56 &  \\
(6,1,1) &  $18.43^\circ$ & $32.31^\circ$ & $\sqrt{56}$ &56 &  2
\end{tabular}
\end{ruledtabular}
\end{table}

Comparing the tables 1 and 2, we observe that  switching from the paradigm  of the symmetrical billiards  to the paradigm of the unsymmetrical billiards  removes the degeneracy of the states to a partial extent. The symmetric quantum billiard possesses  degenerate states which can be termed as sister states with the same path lengths of the closed orbits. The path lengths depend upon the energy eigen value via the sum of the squares of quantum numbers i.e. $(n_{x}^{2}+n_{y}^{2}+n_{z}^{2})$. But the shapes of the closed orbits depend upon how the total energy is distributed amongst the quantum numbers $n_{x},n_{y} ,n_{z}$ i.e. the weight of the $n_{x},n_{y} , n_{z}$. For $n_{x}> n_{y} > n_{z}$, the weight of $n_{x}$ is highest, so the highest proportion of the energy is associated with $n_{x}$. Thus the momentum imparted along the x-axis is maximum. This results in more number of reflections from the x-axis. Mathematically the momentum imparted along the various axis is defined by the angles `$\theta$' and `$\phi$'. Thus the shapes of the closed orbits depend upon the angles `$\theta$' and `$\phi$', which will be discussed in the next section.

\section{Results and Discussions}\label{rd}
Since cubical billiard is a $3D$ system, the visualization of the time evolution of the Gaussian wave packet inside it requires  $4D$ plots which are not easily possible. We have translated these $4D$ plots to a grid in the $3D$. The plots are generated in MATLAB. We have produced  the plots of the probability density  for   $(p,q,r)$  values  $(1,1,1)$ $(3,1,1)$ $(1,3,1)$ and  $(1,1,3)$, for a Gaussian wave packet  centered at $n_0=500$. The $3D$ view of the trajectory for these values  is shown in the figures $(2),(3)$, $(4)$ and $(5)$. The trajectories for other $(p,q,r)$ can be produced following the same procedure. But have not been presented in this work to avoid increasing length of the paper unnecessarily. It is important to select the proper initial co-ordinates $(x_{0},y_{0},z_{0})$ in accordance with the slope that the trajectory will follow. The formulae to calculate these initial co-ordinates, for specific pair of angles $\theta$ and $\phi$  are given below. Here the value of  $x_{0}$ is chosen independently and $|x|=|y|=|z|=1$
\begin{eqnarray}
|y_{0}|&=&|y|-(|x|-|x_{0}|)\tan\theta\nonumber\\
|z_{0}^{'}|&=&\sqrt{(|x|-|x|_{0})^{2}+(|y|-|y|_{0})^{2}}\tan\theta\nonumber\\
|z_{0}|&=&|z|-|z_{0}^{'}
\end{eqnarray}

Figures ~\ref{f7} and ~\ref{f8} are the plots of the probability density showing the dynamics of the Gaussian wave packet in a cubical infinite square well for $(p=1,q=1,r=1)$ at subsequent instants of time. The classical time period of this trajectory is, $T_{cl}^{po} =\sqrt{3}(\approx1.6)\times {\tau_{cl}}$ which is endorsed by plot at  $T_{cl}^{po} = 1.6\times {\tau_{cl}}$. In fig.~\ref{f7} we show the wave packet advancement along the diagonal of the cubical box from $T_{cl}^{po} = 0.1\times {\tau_{cl}}$ to $T_{cl}^{po} = 0.4\times {\tau_{cl}}$. After this time the wave packet collapses due to collision with the infinite potential walls. In fig.~\ref{f8} at about $T_{cl}^{po} = 1.3\times {\tau_{cl}}$, the wave packet revives and starts its backward journey along the same body diagonal.

The figures from ~\ref{f9} to ~\ref{f14} are the plots of the probability density showing the dynamics of the Gaussian wave packet in the cubical billiard for the degenerate energy states $ (113), (131)$ and $(311)$ at subsequent instants of  time. The  classical time period $T_{cl}^{po} = \sqrt{11}\times {\tau_{cl}}$  and length of the trajectory  for these degenerate states is same, but the shapes of the trajectories are different.

Fig.~\ref{f15} shows the plots of the autocorrelation function, $|A(t)|^2$ verses the ratio of classical time period and classical time period of the shortest closed orbit ($T_{cl}^{po}/\tau_{cl}$)  for all the non-degenerate pairs of $(p,q,r)$ i.e. $(1,1,1)$,$(1,1,2)$ ,$(1,3,1)$ and $(1,2,3)$. The time spans from zero to twenty classical time periods. The plots confirm the $T_{cl}^{po}/\tau_{cl}$ values mentioned in the table 1. The decreasing peak maxima validates the spreading of the wave packet for the successive rounds across the billiard.

Fig.~\ref{f16}  plots  the autocorrelation function, $|A(t)|^2$  verses the ratio of revival time period and revival time period of the shortest closed orbit ($T_{rev}^{po}/\tau_{rev}$) for the for all the non-degenerate pairs of $(p,q,r)$ i.e. $(1,1,1)$,$(1,1,2)$ ,$(1,3,1)$ and $(1,2,3)$. The time spans from zero to five revival time periods. The plots confirm the $T_{rev}^{po}/\tau_{rev}$ values mentioned in the table 1. The constant peak maxima for the entire span of the revival time periods indicate the existence of the exact revival pattern in the cubical billiard. The smaller peaks show the presence of fractional revivals.

\section{conclusion}\label{con}
We have visualized the time evolution of the Gaussian wave packet bound inside the cubical billiard and have observed  various closed orbits. The length of the closed orbit depends upon the energy eigen value. Thus the degenerate states of the cubical billiard belong to an identical path length. But the shapes of the closed orbits associated with these degenerate states are different as they depend upon the distribution of energy. The degeneracy originates due the symmetry of cubical billiard, so for a parallelepiped billiard the degeneracy  almost disappears except for few degenerate states that depend upon the ratio of sides of the parallelepiped billiard.
\section*{Acknowledgement}
The work of B.A. is supported by CSIR, grant number 03(1268)/13/EMR-II, India.


\end{document}